\documentclass{article}
\usepackage{arxiv}

\usepackage[utf8]{inputenc} % allow utf-8 input
\usepackage[T1]{fontenc}    % use 8-bit T1 fonts
\usepackage{hyperref}       % hyperlinks
\usepackage{url}            % simple URL typesetting
\usepackage{booktabs}       % professional-quality tables
\usepackage{amsfonts}       % blackboard math symbols
\usepackage{nicefrac}       % compact symbols for 1/2, etc.
\usepackage{microtype}      % microtypography
\usepackage{graphicx}
\usepackage{doi}

\usepackage{multirow}%
\usepackage{amsmath,amssymb}%
\usepackage{amsthm}%
\usepackage{mathrsfs}%
\usepackage{xcolor}%
\usepackage{xspace}%
\usepackage{textcomp}%
\usepackage{manyfoot}%
\usepackage{booktabs}%
\usepackage{algorithm}%
\usepackage{algorithmicx}%
\usepackage{algpseudocode}%
\usepackage{listings}%
\def\pbirth{$p_{birth}$\xspace}
\def\pdeath{$p_{death}$\xspace}
\def\birth{\emph{birth}\xspace}
\def\death{\emph{death}\xspace}
\def\PH{PH\xspace}
\def\PD{PD\xspace}
\def\TDA{TDA\xspace}
\def\PLEX{\emph{Plex Library}\xspace}

\def\DIONYSUS{\emph{Dionysus}\xspace}
\def\PHAT{\emph{Phat}\xspace}

\def\GUDHI{\emph{Gudhi}\xspace}
\def\RIPSER{\emph{Ripser}\xspace}
\def\DIPHA{\emph{DIPHA}\xspace}
\def\PIXH{\emph{PixHomology}\xspace}

\title{A Distributed Approach for Persistent Homology Computation on a Large Scale}

\author{ %\href{https://orcid.org/0000-0000-0000-0000}
{%\includegraphics[scale=0.06]{orcid.pdf}
        \hspace{1mm} Riccardo~Ceccaroni}\\
	Department of Statistical Sciences\\
	Universit\`a di Roma ``La Sapienza''\\
	P.le Aldo Moro 5, Rome 00185, Italy \\
	\texttt{riccardo.ceccaroni@uniroma1.it} \\
	\And
	{%\includegraphics[scale=0.06]{orcid.pdf}
        \hspace{1mm} Lorenzo Di~Rocco}
        \thanks{This work was carried out while occupying the role of visiting student at Bielefeld University, Germany.} \\
	Department of Statistical Sciences\\
	Universit\`a di Roma ``La Sapienza''\\
	P.le Aldo Moro 5, Rome 00185, Italy \\
	\texttt{lorenzo.dirocco@uniroma1.it} \\
	\And
        {%\includegraphics[scale=0.06]{orcid.pdf}
        \hspace{1mm} Umberto Ferraro~Petrillo}\\
	Department of Statistical Sciences\\
	Universit\`a di Roma ``La Sapienza''\\
	P.le Aldo Moro 5, Rome 00185, Italy \\
	\texttt{umberto.ferraro@uniroma1.it} \\
        \And
        {%\includegraphics[scale=0.06]{orcid.pdf}
        \hspace{1mm} Pierpaolo~Brutti}\\
	Department of Statistical Sciences\\
	Universit\`a di Roma ``La Sapienza''\\
	P.le Aldo Moro 5, Rome 00185, Italy \\
	\texttt{pierpaolo.brutti@uniroma1.it} \\
}

\renewcommand{\shorttitle}{\textit{arXiv} Template}

\hypersetup{
pdftitle={A template for the arxiv style},
pdfsubject={q-bio.NC, q-bio.QM},
pdfauthor={David S.~Hippocampus, Elias D.~Striatum},
pdfkeywords={First keyword, Second keyword, More},
}

\begin{document}
\maketitle

\begin{abstract}
Persistent homology (\PH) is a powerful mathematical method to automatically extract relevant insights from images, such as those obtained by high-resolution imaging devices like electron microscopes or new-generation telescopes. However, the application of this method comes at a very high computational cost, that is bound to explode more because new imaging devices generate an ever-growing amount of data.
In this paper we present \PIXH, a novel algorithm for efficiently computing $0$-dimensional \PH on \textsf{2D} images, optimizing memory and processing time. 
By leveraging the Apache Spark framework, we also present a distributed version of our algorithm with several optimized variants,  able to concurrently process large batches of astronomical images.
Finally, we present the results of an experimental analysis showing that our algorithm and its distributed version are efficient in terms of required memory, execution time, and scalability, consistently outperforming existing state-of-the-art \PH computation tools when used to process large datasets. 
\end{abstract}

% keywords can be removed

\section{Introduction}
\label{sec:introduction}

Since the advent of the first electron microscopes, the field of electron microscopy (EM) has undergone significant transformations. Modern microscopy techniques are now achieving near-atomic resolution in the structural analysis of individual proteins and molecular complexes. As a result of this, it is becoming common to generate large digital image files, often above one terabyte in size, in a single data acquisition session \cite{Poger2023}. This advancement poses serious performance challenges when it turns to the analysis of these images. 

A similar problem affects also other application domains. For example, the Vera C Rubin Observatory is a new generation ground-based telescope currently under construction (see \cite{LargeSynopticSurveyTelescope}). When ready, it is expected to generate approximately $15$ terabytes of data per night. Such an impressive amount of data requires very efficient methods to be analyzed \cite{starck2007astronomical}.

In contexts like these, Topological Data Analysis (\TDA) \cite{ComputationalTopologyAnIntroduction} plays a significant role as a tool for the automatic extraction of relevant structural information from large datasets of images. This is especially true for Persistent Homology (\PH) \cite{TopologicalPersistence}, a fundamental component of \TDA, capable of constructing multi-resolution, noise-resilient topological features from a variety of different data clouds  \cite{TopologyandData}. 

Several techniques have been proposed so far for the efficient \PH calculations (see, e.g., \cite{ARoadmap}). However, processing large images within a reasonable time is still impractical. The primary performance bottleneck is the complex computational procedure employed by many of these techniques, i.e., the filtration of simplicial and cubical complexes \cite{kaczynski2004computational}. The execution cost of this procedure increases exponentially with the size of the input data. For this reason, innovative algorithms and software solutions that can efficiently handle vast datasets of very large images are required. 

%of topological features and their robustness across various spatial resolutions \cite{TopologyandData}. 
%\PH has drawn attention in the field of data analysis, mainly because it excels in extracting topological information that is resilient to noise.
%
%
%
%{\red These methodologies find active application in image analysis, employed for tasks such as segmentation \cite{segmtoppers, topopreservsegm}. This is due to the close association between \PH and the identification of connected components.}
%
%
%For example, keeping our focus on astronomy, this family of techniques has shown great potential in the identification and study of massive galaxy clusters within a target image (see \cite{tis}).

%More in particular, the $0$-dimensional \PH is strictly related to the number of connected components and it is typically visualized on a Persistence Diagram (\PD) with its associated \birth and \death coordinates \cite{StructureandStability}. In the analysis of deep-field astronomical images, \birth refers to the maximum intensity at the center of the galaxy, while \death indicates the minimum intensity at the galaxy's edge.

%Since astronomical images can easily reach very large sizes (e.g., more than $10, 000 \times 10, 000$ pixels), to process using the current methods can be challenging. 

In this work, we propose \PIXH, a novel algorithm for computing the particular case of $0$-dimensional \PH on digital images, that speed up the filtering of a simplicial complex. Our approach
offers a substantial reduction in memory usage compared to existing methods, like the one employed by the state-of-the-art \RIPSER package, when applied to $0$-dimensional \PH computation. 
We also introduce a software pipeline for using \PIXH on a distributed system, to compute $0$-dimensional PH on large batches of images. 

We evaluate the performance of our algorithm and its distributed version using a reference dataset of images and show its efficiency compared to existing methods and software, making it the fastest algorithm available nowadays for $0$-dimensional \PH computation, both in its sequential and distributed versions.

\paragraph{Organization of the paper}
In Section \ref{sec:background}, we provide a short introduction to the theoretical concepts behind the \PH computation problem. Then, in Section \ref{sec:relatedwork} we review the existing literature on computational methods and software tools for efficient \PH evaluation. In Section \ref{sec:spark}, we briefly
%
%invertire MapReduce-Spark
describe the MapReduce distributed computing paradigm and its implementing framework, Apache Spark. Following this, in Section \ref{sec:contribution}, we present our novel algorithm for efficiently computing $0$-dimensional \PH and its distributed version, together with several variants we developed to improve upon its original performance. 
In Section \ref{sec:experiment} we report the results of a thorough experimental analysis designed to assess the performance of our algorithm also in comparison with other existing state-of-the-art tools and methods for \PH computation. Finally, some concluding remarks are given in Section \ref{sec:conclusions}. 

%%%%%%%%%%%%%%%%%%%%%%%%%%%%%%%%
%%%% Theoretical Background %%%%
%%%% (da posizionare)       %%%% 
\section{Theoretical Background}
\label{sec:background}
\subsection{Simplicial and Cubical Complexes}
Simplicial and cubical complexes are the fundamental building blocks of computational topology to fully describe topological spaces. Simplicial complexes consist of simplices, such as vertices, edges, and triangles. In general, a $d$-simplex represents the convex hull of $d+1$ points, and each subset of these $d+1$ points forms a face of this $d$-simplex. A collection of simplices, denoted by $K$, represents a simplicial complex if it satisfies two conditions: all faces of a simplex in $K$ also belong to $K$, and the intersection of any two simplices in $K$ is either empty or a common face.

For cubic complexes, an elementary interval can be described as a unit interval $[k, k + 1]$ or as a degenerate interval $[k, k]$. For a $d$-dimensional space, a cube is the product of $d$ elementary intervals, denoted $\prod_{i=1}^{d} I_i$. The dimension of a cube is determined by the number of non-degenerate intervals in this product. In particular, $0$-cubes, $1$-cubes, $2$-cubes, and $3$-cubes correspond to vertices, edges, squares, and 3D cubes, respectively. When comparing two cubes $a$ and $b$ in $\mathbb{R}^{d}$, $a$ is considered to be the face of $b$ only if $a$ is contained in $b$. A cubic complex of dimension $d$ consists of cubes of dimensions at most $d$. Similar to a simplicial complex, it must be closed under operations with faces and intersections.

For an in-depth discussion of these topics we refer the interested reader to  \cite{GeometricandTopologicalInference,ComputationalTopologyAnIntroduction}.

\subsection{Persistent Homology}
\PH is a fundamental concept in \TDA specifically focusing on $Z^{2}$ homology (see \cite{ElementTopology,ComputationalTopologyAnIntroduction} for a thorough introduction to this topic).

In the context of \PH, we start with a topological space $X$ and a filtering function $f : X \rightarrow R$. This method examines the homological transformations of the sublevel sets, denoted as $X_t = f^{-1}(-\infty, t]$. The algorithm captures the inception and extinction times of the homology classes as the subsets evolve from $X^{-\infty}$ to $X^{+\infty}$. For example, it identifies components as $0$-dimensional homology classes, tunnels as $1$-dimensional classes, voids as $2$-dimensional classes, and so on. \emph{Birth} implies the emergence of a homology class, while \death implies its trivialization or amalgamation with another class that emerged earlier. The persistence or lifetime of a class represents the time difference between its death and birth. Homology classes with greater persistence provide information about the global structure of the space $X$, which is affected by the function $f$.

\begin{figure}[ht]
 \centering
  \includegraphics[width=10cm]{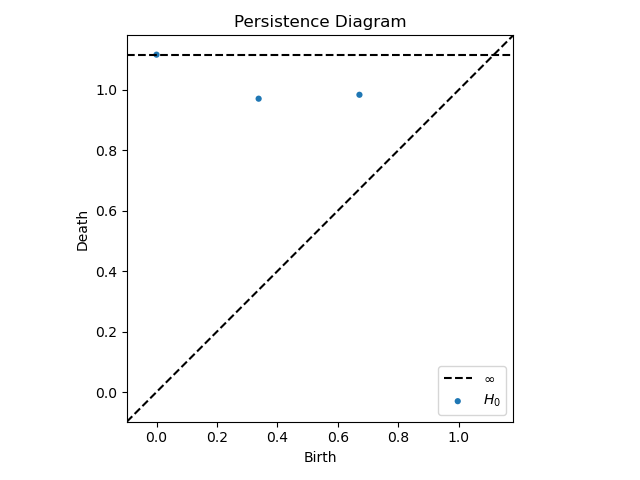}
  \caption{A point $(x,y)$ in the \PD indicates a topological feature of dimension $0$ ($H_0$) born at $x$ and that persists until $y$. We call $x$ the {\pbirth} and $y$ the \pdeath. By definition, all points should lie above the diagonal. The horizontal dashed line represents infinity.}
  \label{fig:pd}
\end{figure}

A common method for visualizing persistence is a Persistence Diagram (\PD), shown in Figure \ref{fig:pd}, consisting of points on a two-dimensional plane, each corresponding to a \PH class. These points are defined by their birth and death times.

A key reason for using persistence is the stability theorem \cite{StabilityofPersistenceDiagrams}: for any two filtering functions $f$ and $g$, the difference in their persistence is always bounded by the $L^{\infty}$norm of their dissimilarity:
$$
|| f - g ||_{\infty} := \max_{x \in X} |f(x) - g(x)|.
$$
This ensures that persistence serves as a distinctive signature. If two persistence outputs differ, it means that the functions are different.

\subsection{Computation of Persistence}
The original algorithm for computing persistence \cite{TopologicalPersistence} operates in cubic time relative to the size of the complex. This approach requires preprocessing of the data. In the case of images, the function $f$ is defined for all pixels. These values are first interpreted as the values of the vertices of the complex. Then, the complex filtration is calculated and a sorted boundary matrix is created.

During filtration, the process entails adding cells with increasing values to the complex one by one. To achieve this, an algorithm for building the filtration extends the function to all cells within the complex by assigning each cell the maximum value among its vertices. Then, all cells are sorted in ascending order according to the function value. As a result, each cell is added to the filtration according to all of its faces, creating a sequence of cells known as \textit{lower-star filtration}. This ordering of cells allows the creation of a sorted boundary matrix.

In the reduction phase, the algorithm performs column reductions on the sorted boundary matrix, proceeding from left to right. Each new column is reduced by adding it to already reduced columns, to maximize the lowest non-zero entry. The final reduced matrix contains all the information about \PH.%%%%%%%%%%%%%%%%%%%%%%%%%%%%%%%%

\section{Related Work}
\label{sec:relatedwork}

Several methodological and software contributions have been proposed so far to support the efficient calculation of {\PH}.  One of the first software tools to be proposed for this purpose is the \PLEX, developed by the Computational Topology Group at Stanford University \cite{adams2014javaplex}. 
%
%\DIONYSUS \cite{dionysus} has been instead the first software package to implement the dual algorithm \cite{de2011dualities, de2009persistent}: it was the first library to implement computation of the bottleneck and Wasserstein distances \cite{Villani2009}. 
\DIONYSUS \cite{dionysus} has been instead the first software package to implement the dual algorithm. Starting from the observation that cohomology groups are usually faster to compute, this algorithm reformulates the problem of homology group computations into a cohomology group computation problem (for more info see \cite{de2011dualities, de2009persistent}).
%\DIONYSUS was also the first library to implement the computation  of the bottleneck and Wasserstein distances \cite{Villani2009}.

\PHAT \cite{PHAT} is the first software to implement a matrix-reduction algorithm that can be executed in parallel, to accelerate the analysis of large datasets. 
%\PERSEUS was developed to implement Morse-theoretic reductions \cite{perseus}. 
\GUDHI \cite{maria2014gudhi} implements a comprehensive library offering functionalities from basic to advanced {\PH}, including new data structures for simplicial complexes and the boundary matrix.
Finally, \RIPSER \cite{Bauer2021Ripser} is considered the gold standard solution in this field, thanks to its versatility and efficiency. It uses several optimizations and shortcuts to speed up the computation of \PH in all dimensions and has demonstrated superior performance to other software tools in terms of both speed and memory efficiency \cite{ARoadmap}.

Indeed, the analysis of very large datasets can be computationally prohibitive, even for efficient tools like \RIPSER. A natural solution to this problem is distributed computing. \DIPHA \cite{DIPHA} has been so far the first software we know of that implements \PH computation with distributed computing, enabling efficient processing of large dimensional data. It works by efficiently partitioning the \PH computation problem into subprocesses to be concurrently run on the nodes of a distributed system. This allows \DIPHA to compute PH on much larger instances than would be possible on a single machine. Moreover, the performance speed-up granted by parallelism introduced by DIPHA at least compensates for the overhead caused by communication between nodes.

Recently, alternative high-performance software solutions like CubicalRipser \cite{cubicalripser} and Cubicle \cite{cubicle} have been introduced for the PH computation on images.

\section{Apache Spark}
\label{sec:spark}

Apache Spark \cite{spark} is one of the most popular engines for large-scale data processing and is based on RDDs (Resilient Distributed Datasets) and DataFrames. These are distributed memory abstractions that allow programmers to perform in-memory computations on large clusters in a fault-tolerant manner. The former are collection of key-value pairs to be processed by means of distributed transformations. The latter are table-like collections provided with the Spark SQL module, which optimizes structured data processing by introducing SQL-like logic in a distributed context. 

%Dataset and DataFrame provide a representation of the distributed dataset as a table with a custom schema, even if the rows are scattered across worker nodes.

\begin{figure}
 \centering
  \includegraphics[width=10cm,height=5cm]{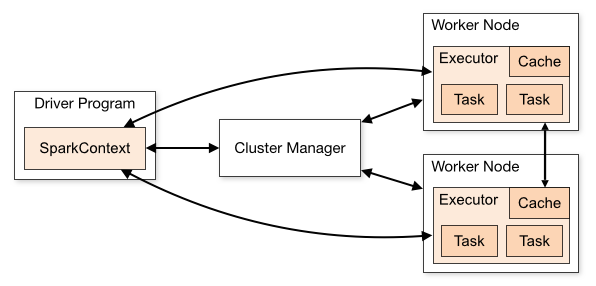}
  \caption{Apache Spark architecture. Example for a reference installation featuring two worker nodes and one driver application. Each worker node in this figure runs one executor process and two tasks. The overall distributed execution is orchestrated by a cluster manager.}
  \label{fig:sparkarchitecture}
\end{figure}

The physical architecture of a Spark cluster, shown in Figure \ref{fig:sparkarchitecture}, is characterized by a master node that oversees a set of \textit{worker} nodes via daemon processes. Data is typically distributed among the worker nodes and MapReduce is also supported.

A Spark application communicates with a \emph{Cluster Manager}, which is the process that manages the computing and storage resources of the cluster. It includes a \textit{driver} process and a set of \textit{executor} processes. The driver process communicates with the Cluster Manager to learn where the data is located and what physical computing resources are available. Then, in each worker node, a set of parallel executor processes is activated according to the number of threads. For example, a cluster with three worker nodes, each with two threads, means a potential number of six parallel executor processes.

\subsection{The MapReduce paradigm}
The computations within Spark are formulated according to MapReduce, a programming paradigm that allows massive scalability (\cite{mapreduce}). MapReduce is composed of two tasks: mapping and reducing. The mapping phase transforms a dataset into another form with elements organized into key-value pairs. Subsequently, the reduction process uses the output of a map as input and combines those data tuples into a smaller set of tuples. As the name suggests, reduction always follows mapping . Consequently,  assuming the input data set is organized as a set of key-value pairs, it is initially distributed among the worker nodes of a cluster. Then, batches of key-value pairs are processed in parallel by concurrent executor processes on the worker nodes where this data is found.  Reduce functions require a preliminary step to group on the same node all pairs having the same key (\textit{shuffle} operation). When working with large data sets, this preliminary step makes the Reduce function potentially expensive from a computational perspective.

%MapReduce transformations in Apache Spark are issued on \textit{RDDs} (Resilient Distributed Datasets), which are abstract data structures representing all key-value pairs to be processed.
%Spark provides structured representations of the partitioned data: \textit{Dataset} and \textit{DataFrames}. These distributed data structures are table-like collections provided with the Spark SQL module, which optimizes structured data processing by introducing SQL-like logic in a distributed context. Dataset and DataFrame provide a representation of the distributed dataset as a table with a custom schema, even if the rows are scattered across worker nodes.

\subsection{Fault-tolerant applications}
Spark supports various fault tolerance strategies including checkpointing, task replication, and error handling, all of which contribute to the reliability and robustness of the computing process. For instance, checkpointing allows Spark to periodically save the state of distributed data structures to resilient storage, enabling recovery from failures without recomputing from scratch, while task replication ensures that tasks are rerun on different nodes in case of failures \cite{fault_tolerant_abstraction, sparkdocumentation}.

\section{Our Contribution}
\label{sec:contribution}
%\red

%To be completely rewritten

%A natural way to accelerate PH computation  would be to resort to parallel computing or to distributed computing. However, unexpectedly, only a very small number of PH software tools are today capable of exploiting parallel computing and, as far as we know, none of them can be used to analyze images (see Section \ref{sec:relatedwork}).

%Distributing the PH computation of astronomical images is very challenging. Splitting the image into patches to process in parallel may introduce a heavy data traffic, burdening the time performances. This happens because computational units processing adjacent patches may have to frequently communicate to detect the birth and death coordinates falling in their patches. We introduce here a novel distributed pipeline, based on the MapReduce paradigm, to compute the PH of a batch of astronomical images, exploiting the resources of a distributed system to overcome the memory limits of a single workstation and speed up the elaborations.

In this section, we introduce a novel algorithm for computing \PH on \textsf{2D} images: the \PIXH algorithm \footnote{The source code of \PIXH is available at \url{https://github.com/riccardoc95/Sparksistence}}. Our algorithm comes with two relevant advantages concerning the existing literature. First, it offers a substantial reduction in memory usage compared to existing methods like the \emph{lower\_star\_img} function of \RIPSER package, when applied to $0$-dimensional \PH computation. Second, it has been conceived to process large batches of images in parallel using a distributed system in a more efficient way than other distributed systems (i.e., \DIPHA).

The first goal has been achieved by overcoming a relevant performance bottleneck existing in traditional general filtration \PH algorithms, i.e., the computation of adjacency matrices. Being purposely designed to deal with the particular case of $0$-dimensional \PH computation, our algorithm can avoid all the computational burden of constructing and analyzing these matrices while dramatically reducing the overall amount of memory required for its execution.

%designing a novel \PH algorithm that does not require adjacency matrices, a key component of conventional \PH algorithms. This allows us to dramatically reduce the amount of memory required for executing the algorithm and avoid the significant computational burden of constructing and analyzing these matrices. 

%approach enabled a memory consumption that scales linearly with the number of pixels in the image. 

The second goal has been reached by using the MapReduce paradigm to develop a distributed \PH pipeline based on our algorithm. This allows the execution of the algorithm concurrently on very large batches of images, in an efficient and scalable way.

\medskip

% In the end of this section, we introduce a novel distributed pipeline, based on the MapReduce paradigm. 

\subsection{The \PIXH Algorithm}
The algorithm we propose, here called \PIXH, has been designed to process efficiently very large images as input while yielding zero-dimensional \PH as its output. Specifically, the algorithm will provide the \birth and \death values of each object within the image, along with their pixel coordinates.

\begin{figure*}[t]
  \centering
  \includegraphics[width=\textwidth,height=4cm]{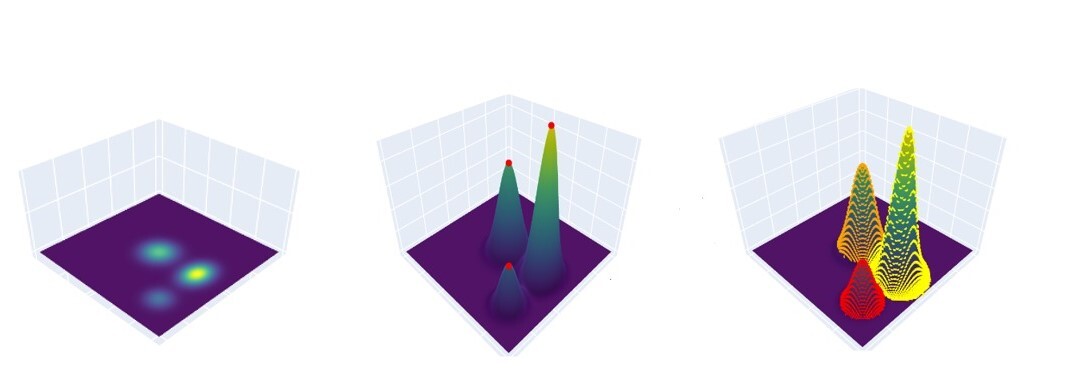}
  \caption{ The PH calculation using \PIXH on an image containing three components defined by Gaussian functions. Initially, each pixel is linked to its neighbor with the highest value, and \PIXH detects relative maxima as \birth values. Subsequently, all the minimum or saddle points are located. The first value of these points that connects the two components represents the \death value of the component with the lower \birth value. Finally, the process ends with identifying the absolute minimum in the image, which serves as the ultimate death point associated with the component relative to the absolute maximum.
}
  \label{fig:methods}
\end{figure*}

The straightforward implementation and the computational efficiency result from constraining the application to \textsf{2D} images with specific characteristics.
\noindent The algorithm initiates by linking each pixel to its highest value neighbor among the $8$ surrounding pixels. This process enables the division of the image into connected components, which are later united to generate the points on the \PD. A crucial condition for the proper operation of \PIXH is that the pixel containing a local maximum value must not have any neighboring pixels with the same value among its $8$ neighbors.
While the application domains of this algorithm may appear limited at this stage, it's important to note for example that whenever Gaussian noise is added to a signal in an image, the image satisfies the necessary conditions for the application of \PIXH.

One of the key points of our proposal, as visible in Algorithm \ref{alg:pixh_alg}, is the usage of a $maxpool2d$ function with a kernel size of $3$, a stride of $1$, and a padding of $1$. This function yields an image of the same initial dimensions, with each pixel's value being set to the maximum among the pixel and its eight neighboring pixels.
The $arg\text{-}maxpool2d$ function employs identical parameters as the $maxpool2d$ function. However, instead of returning the pixel values with the highest value, it returns the indices corresponding to these pixels.

\noindent Given an input image $I$, the algorithm we propose requires the following steps.

\begin{itemize}
    \item{\em Step 1: Identification of the concave components}.

This step is about the identification of the concave components within the image $I$. 

To accomplish this, we use the $arg\text{-}maxpool2d$ function to compute a new image $M$ of the same initial dimensions of $I$, with each pixel's value being set to the maximum among the pixel and its eight neighboring pixels. Then $M$ is processed in a loop. 

In each iteration, every element $x \in M$ is replaced by the value $M[x]$, which corresponds to the value located at position $x$ of $M$. This process ends when $x$ is equal to $M[x]$ for every element $x \in M$.

%To accomplish this, we compute, for each pixel, the index of the pixel with the highest value among its eight neighboring pixels, using the $arg\text{-}maxpool2d(x)$ operation. 

%Then $M$ is processed in a loop. In each iteration, every element $x \in M$ is replaced by the value $M[x]$, which corresponds to the value located at position $x$ of $M$. This process ends when $x$ is equal to $M[x]$ for every element $x \in M$. %Because the extreme value theorem guarantees the existence of absolute maxima, we can be confident that this process ends within a finite number of steps.

%Currently, the detected elements are labeled with the index of their neighbor relative maximum. 
%\red 
%are you talking about the labeled image $M$?

%in the title we talk about birth points and components, while these terms are missing in the following text. You should be consistent with the terminology you use
%\black
%We gather all the indices of relative maxima into an array $pbirth$. Then, we assign sequential increasing numbers to the zones, starting from the one with the smallest relative maximum to the one with the largest one.

\item{\em Step 2: detection of birth points and re-indexing of components}

At this point, all the identified elements in $M$ are marked with the index of their corresponding relative maximum neighbor found in $I$. 
The unique values of M are the position of the relative maximums that are stored in an array labeled \pbirth, i.e. the birth points. A separate \birth array records the values of $I$ at these \pbirth points. We sort \birth and \pbirth array so that \birth array is in descending order.
Subsequently, we update the values in $M$ with positions corresponding to the values in the \pbirth array. This process assigns incremental numbers to the components of $I$, starting with the component that has the smallest relative maximum and ending with the component that has the largest relative maximum.

%\red 
%we are assuming that the reader is familiar about notions like edges and dead points. This is likely to be true but, anyway, it would be better to spend more words about them or to reference the section where these notions are introduced and explained
%\black

%After have partitioned the image into distinct components, we detect the edges, and within the found edges we later locate the dead points 
%\red I would suggest macros for terms like pdeath, pbirth, etc
%\black
%\pdeath. To detect the edges, we calculate the $maxpool2d(M)$ and $-maxpool2d(-M)$. The area of our interest will be the region where these two results differ. We provide the indices that specify this region within array $B$.

\item{\em Step 3: edge points detection}

Upon partitioning the image into distinct components, we calculate $maxpool2d(M)$ and $-maxpool2d(-M)$. The region where these two outcomes differ represents the edges of the components within the image $I$. It is essential to note that matrix $M$ comprises integer values, each signifying a unique component in the image. These component values are not arbitrarily assigned; rather, they follow the order of the relative maxima present in $I$.

An array $B$ that contains the indices of all the edges of the components is generated.

%After splitting the image into different components, we calculate the $maxpool2d(M)$ and $-maxpool2d(-M)$. The area in which these two results differ represents the edges of the objects in the image $I$.
%We specify the indices that specify this region in the array $B$.

\item{\em Step 4: distillation}

By definition, death points are located along the edges of the components. To connect two neighboring components, they must be either relative minimum points or saddle points. In this step, we verify if the points with an index in $B$ are minimum or saddle points. If these criteria are unsatisfied,  the index is removed from $B$.
We characterize a minimum point as a pixel with the lowest value in comparison to its $8$ neighboring pixels. In contrast, a saddle point is a pixel that serves as a minimum along one axis and a maximum along the other axis, always about its $8$ neighboring pixels.

%By definition, death points are located along the aforementioned regions. To connect two neighboring objects, they must be either relative minimum points or saddle points. In this step, points that do not meet these criteria are removed from $B$.

\item{\em Step 5: dead points identification and partition merging}

We sort $B$ in descending order to maintain the chronological sequence of partition merges in $M$. Each point $x$ in $B$ that is adjacent to two partitions triggers their merger. We call $x$ the \pdeath\/ point for the lesser-indexed partition. Merger history is captured in vector $C$, which stores the new index of each partition after the merger.

To further improve the efficiency of the algorithm, we restrict the changes to the eight pixels around the point $x$, rather than the entire array $M$.

\item{\em Step 6: \PD construction}

In this step, we create a  \PD, to associate each birth point with its respective death point within the same partition. We then extract the $birth$ and $death$ values for the \pbirth and \pdeath points from $I$ and aggregate them into a $DGM$ matrix.

\end{itemize}

When compared with existing literature, \PH achieves minimal memory usage and efficient execution times by eliminating the classical \textit{adjacency matrix} in step $1$ and accelerating filtration in step $5$ by avoiding pixel-by-pixel control.  In terms of computational complexity, each operation in  Algorithm \ref{alg:pixh_alg} incurs a cost of $O(n)$, where $n$ is the number of pixels in image $I$, except for the while loop. In a highly improbable scenario where there is only one component in the image with \birth value in the last pixel of $I$, the while loop concludes after $n(n-1)$ operations.

%The algorithm's advantages are in step $1$ and in step $4$, where they eliminate the classical \textit{lower-star filtration} and accelerate filtration in the step $5$ by avoiding pixel-by-pixel control. These optimizations result in \textit{PixHomology} exhibiting minimal memory usage and efficient execution times.

\begin{algorithm}[ht]
\caption{Outline of the \PIXH algorithm. It assumes the availability of the $unique$ function, which is used to extract the unique elements from an array, the $reindex$ function, which is used to reset the indexes of the components so that the highest index corresponds to the component containing the pixel with the greatest value, and the maxpool2d and arg-maxpool2d functions which return respectively the output of maxpool operation on \textsf{2D} images and the indexes of this operation. Additionally, the $distillation$ function, detailed in the step with the same name, identifies and removes unnecessary pixel indices in subsequent steps.}
\label{alg:pixh_alg}
\textbf{Input}: Image $I$\\
\textbf{Output}: \PD, the matrix $dgm$ with birth and death values
\begin{algorithmic}[1]
\State $M = \text{arg-maxpool2d}(I)$
\While{$\exists \; x \in M: x != M[x] \;$}
\State $x == M[x] \; \forall x \in M$
\EndWhile
\State $\text{\pbirth} = unique(M)$, $M = reindex(M)$
\State $B = \text{where(maxpool2d}(M) != -\text{maxpool2d}(-M)\text{)}$
\State $B = distillation(B)$
\State the elements within B are arranged in a descending order%$B = sort(B)$
\ForAll{$x$ in $B$}:
    \If{$x$ is border of two partitions $P_1$ and $P_2$}
        \State $x$ is added in \pdeath
        \State $C[x] = min(P_1,P_2)$
    \EndIf
\EndFor
\State $dgm = (I[\text{\pbirth}], I[\text{\pdeath}])$
\end{algorithmic}
\end{algorithm}

\subsection{The Distributed \PIXH Pipeline}
\label{sec:sparkpipeline}

Tools like \DIPHA overcome the heavy memory and computational requirements of many PH algorithms by distributing the computation across multiple nodes. However, this approach presents a significant challenge: dividing the image into patches for parallel processing reduces the memory required per node, but it also leads to substantial data traffic between nodes, as the computational units handling adjacent patches must communicate frequently to detect \birth and \death coordinates within their respective regions.

\begin{figure}[ht]
  \centering
  \includegraphics[width=10cm]{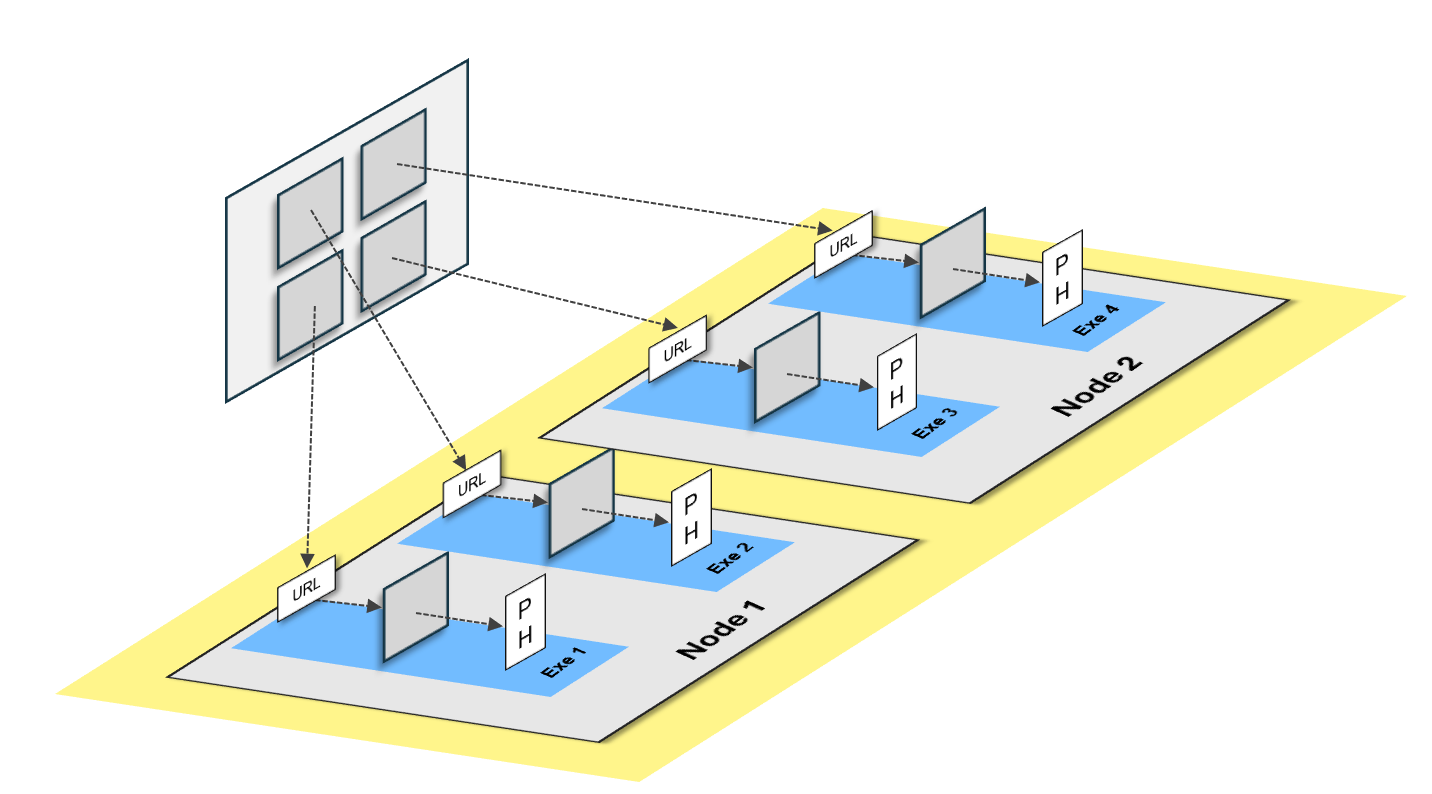}
  \caption{An overview of the distributed workflow of \PIXH on a Spark cluster involving four executor processes scattered across two computing nodes. After partitioning the URLs of the images across the various executors, each executor performs two map operations. The former operation loads the image into memory, while the latter performs the PixHomology algorithm to compute the 0-dimensional PH. }
  \label{fig:workflow}
\end{figure}

%The \DIPHA software successfully employs this technique to distribute homology calculations for a single image. It optimizes memory usage and distributes the workload efficiently through an optimized representation of cubical complexes provided by the CubeMap function \cite{DIPHA}.

%On the other hand, \PIXH proactively reduces memory usage.
%This approach eliminates the need to segment the PH calculation into sub-processes, allowing us to leverage computing resources for the concurrent processing of multiple images.

Given the expected smaller memory footprint of our algorithm, we were able to adopt a different solution: concurrently processing multiple images at once using the different computational units of a distributed system. Based on this idea, we developed a simple distributed pipeline for our \PH algorithm using the Apache Spark framework. The choice of this technology over other distributed computing frameworks has been motivated by its inherent scalability and by its ability to better operate on cloud-based big data processing infrastructures.

The proposed pipeline, shown in Figure \ref{fig:workflow}, consists of two Spark distributed transformations, followed by one Spark distributed action to collect the computation results. 
The first distributed transformation, implemented as a \textit{map} operation, is named $load\_preprocess\_image()$. It is used by each Spark executor to load each input image in memory and prepare auxiliary data structures used by the algorithm for subsequent steps. Images are retained in memory using \textsf{2D} array representations.

The second distributed transformation, also implemented as a \textit{map} operation, is named $process\_image()$. It is used by each executor to apply the \PH algorithm to each of the arrays loaded in the previous step, yielding $0$-dimensional \PH.

% Finally, the \textit{collect} distributed action is used to gather the results on the master node.

\subsubsection{Variants}
\label{subsubsec:variants}
Once ready, we performed a preliminary experimental analysis targeting our distributed algorithm,  to identify hotspots and address possible performance bottlenecks. The insights gained from this analysis were used for the development of the following more efficient variants.

%{\red Q: Optimization}
\begin{itemize}
    \item {\textbf{Variant 1: reducing images loading time.}} 
    
    A simple yet effective way to handle input data in Spark is to load the entire dataset in the memory of the driver application and then convert it into a distributed RDD representation using the \texttt{parallelize()} method. This approach fails with huge datasets both because of the large amount of memory required by the driver application to initially store the datasets, and because of the long execution times needed to first load in memory the dataset and then distribute it across the cluster.
    
    We developed an alternative approach where each executor loads the images to be processed on its own, either from the local disk or from a remote web server. This solution alleviates the pressure on the driver application and reduces loading times. From the technical viewpoint, this solution was implemented by avoiding the need for the driver to load any images. Instead, we modified the original $load\_preprocess\_image()$ distributed transformation to instruct executors about the physical location of the images to load and process. This variant is much faster than the original one, so we will use it in all the following experiments. Notice that our improved approach is still vulnerable to the case where the size of the images to be processed exceeds the amount of memory available to a single executor. Such a problem could be circumvented by reducing the number of executors to use, so as to increase the amount of memory available to each of them.

    %{\red \textbf{Q: Memory management}\\
    %We observe that although \PIXH requires significantly less memory than other existing methods, as RDDs consist solely of image pathnames, it makes the process vulnerable: the proposed algorithm lacks a solution if an individual image exceeds the memory capacity of a single worker. However, Spark automatically manages the broader scenario wherein the collective size of all images surpasses available memory. This is facilitated by Spark's mechanism whereby any oversized partition of an RDD allocated to a node is seamlessly offloaded onto disk storage \cite{sparkdocumentation}.}

    \item {\textbf{Variant 2: reducing the amount of pixels to process.}}
    
    %{\red da rimuovere (?!?)}

    %During the analysis of digital images, the processing of every single pixel may  not be imperative. This can be due to the existence of noise in the pixels, that can obscure the the presence or absence of objects. Therefore, a threshold $t$ is typically established, below which pixels are designated as background. Some of the techniques for setting this threshold are described in \cite{starck2007astronomical}.
    In the analysis of digital images, processing every individual pixel might not be imperative. Hence, there could be situations where it proves beneficial to set some sort of threshold value $t$, below which pixels do not necessitate examination.

    Applying this technique can be of significant help when analyzing large batches of high-resolution images, as it can significantly reduce the computational burden. In our case, we consider a very simple preprocessing method, where we assign a threshold to each image within the dataset. This threshold is acquired with each image and then applied to the image itself using the $load\_preprocess\_image()$ function in each experiment. Consequently, when a new image is loaded, all pixels with a value under this threshold are identified as \textit{background pixels} and excluded from the subsequent analysis. Notice that we do not aim to find an optimal thresholding strategy, but rather to assess the potential performance gain achievable by applying this technique. 

%    In the implementation of \PIXH, background pixels are identified and excluded during the $distillation$ step. 

%This allows us to pre-evaluate the expected execution times, which is subsequently used to balancing the computational workload: images containing a higher number of background pixels undergo faster processing.

%In its implementation, \PIXH identifies these background pixels and excludes them from the $distillation$ step, which results in an acceleration of our pipeline. In fact, since \PIXH handles images with a greater proportion of background pixels more rapidly, we can make a preliminary assessment of execution times in advance.

%We then assign a threshold $t$ to each image in the dataset, below which object detection is not possible. This threshold is used as a preprocessing step before analyzing the image. The techniques for setting this threshold are described in \cite{starck2007astronomical}.

    \item {\textbf{Variant 3: improve the workload distribution.}}

    When a distributed data structure is created in Spark, it is automatically partitioned into a number of partitions based on factors such as the number of executors and the size of the data. The developer can override these settings and specify the number of partitions to use.

    Partitioning is important because partitions are the basic units of parallelism in Spark. A good partitioning strategy will ensure that all executors in a distributed system have approximately the same workload, resulting in shorter execution times.

    In our particular case,  the processing time of each image is relatively long. Provided that each executor is fed with approximately the same number of partitions, this does not necessarily guarantee a balanced workload, as the processing time for each image may vary depending on several factors. Moreover, because the amount of data stored in each partition is very small (i.e., the address where each image is stored) and the time spent transmitting it over the network is negligible, there is no need to ensure the locality of data in the worker nodes. To account for these factors, we tried several partitioning strategies for our pipeline.

    \begin{itemize}
        \item \textit{Strategy 1: partitioning by the number of executors}

        Let $n$ be the number of input images and $m$ be the number of executors available on the distributed system. We partition the image dataset into $m$ partitions, one for each executor. To do so, we first create a list of $n$ integers, where each integer represents the index of an image in the dataset. We then shuffle the list of integers and divide it into $m$ partitions, each containing $n/m$ integers. Finally, we assign each partition to an executor.

        This partitioning strategy ensures that all executors have approximately the same workload, as each partition contains the same number of images. It also minimizes the amount of data that needs to be transmitted over the network, as each executor processes the images in its assigned partition.
        However, it can perform poorly with datasets that have a skewed image complexity distribution. If an executor is assigned a partition with images requiring very long processing times, its execution time will dominate the overall algorithm execution time, making it a \textit{straggler}.

        \item \textit{Strategy 2: partitioning by the number of images}

        Let $n$ be the number of input images and $m$ be the number of executors available on the distributed system. We partition the image dataset into $n$ partitions, one for each image. Then, we let Spark assign partitions to executors using the default partitioning strategy. 

        As before, this partitioning strategy ensures that all executors have approximately the same workload, as each executor is initially assigned approximately the same number of partitions. Differently from the previous case, we expect this strategy to be more effective in handling datasets that have a skewed image complexity distribution. If an executor is assigned a partition with an image requiring processing times much longer than other partitions, Spark may automatically reassign the remaining partitions to other executors, thus mitigating the effects of that straggler on the overall algorithm execution time. 

        \item \textit{Strategy 3: overriding standard partitioning rules}

        %{\red da rimuovere (?!?)}

        We expect strategy 2 to help mitigate the effects of stragglers, but still under the possibly wrong assumption that all images have approximately the same processing time. To overcome this problem, we introduce a third partitioning strategy. We estimate the computational cost for the analysis of input images and then use the Lasting Processing Time (LPT)  rule \cite{lpt} to schedule the analysis of input images on all executors to ensure a uniformly distributed workload. 
        
        LPT provides a heuristic solution to the NP-hard scheduling problem with identical parallel machines and no preemption. It has already been used in literature to minimize the completion time (also referred to as \emph{makespan}) of Spark jobs, as shown in \cite{scheduling}.

        The goal consists of distributing $n$ jobs, characterized by specific processing times $\{p_{j},j=1,\dots,n\}$, among a set of $m$ executors operating in parallel to minimize the maximum makespan.  Basically, the LPT rule sorts the jobs in a non-increasing order according to their processing times and, iteratively, assigns a job to the machine that currently has the minimum completion time. 
        
        In our case, each job corresponds to the PH calculation on a single image, and we estimate the processing time as proportional to the number of pixels that have been identified as non-background during the preprocessing phase (see Variant 2).

\end{itemize}

\end{itemize}

%Firstly, our pipeline flattens the input image and broadcasts a copy of it to each node of the cluster, so that it will be accessible to all computational units of that without further copies. Then, a distributed data structure storing the coordinates of each pixels for the input image is generated.  

%Next, using a map function, each  pixel is processed in parallel to search for its closest birth coordinates (also called \emph{root coordinates}). Notice that this happens without any further communication step because each computational unit can advantage of the local copy of the entire image. 

%Finally, a reduce function merges all the pixels associated with the same root coordinates to detect the corresponding death coordinates by computing the pixel associated with the minimum intensity level. 

%The output of the pipeline consists of a data structure reporting, for each birth point, its corresponding death point.\\

% The Spark pipeline starts creating an RDD that contains a list of the file paths of the dataset images and the threshold level $t$ for each image (see section \ref{sec:dataset}).

\section{Experimental Analysis}
\label{sec:experiment}

In this section, we present the results of an experimental analysis done to assess the performance of our algorithm and its variants, according to several metrics. We also compare its performance with the state-of-the-art software in this field. While recent software \cite{cubicalripser, cubicle} has demonstrated outstanding performance in computing \PH on images, this section will focus on comparing results with the \RIPSER package. This decision is based on potential differences in output between software using cubical complexes and those analyzing simplicial complexes. However, a comparison will also be conducted with \DIPHA, the only software implementing distributed calculation, despite its use of cubical complexes.

%In this section, we provide an example utilizing images from an astronomical dataset.
%In the case of these images, we can expect that the \birth points will align with the centers of the stars in the image. Conversely, the \death points will occur when the brightness resulting from two stars joins or when the brightness of a star fades into the background.

%We performed an experimental analysis to assess the output of \PIXH in comparison to established software for \PH calculation. Additionally, we evaluated the execution time and memory usage. We compared our solution with \RIPSER, the gold standard software. Subsequently, given our approach's incorporation of distributed computing, we conducted a comparative analysis of our pipeline with \DIPHA, focusing on execution time.

\subsection{Computing Environment}

Our experiments were conducted on the Terastat HPC infrastructure of $24$ nodes, where each node is equipped with $128$ computing units and a maximum of $2$ GB RAM of memory per computing unit (see \cite{TeraStat} for more details).
Experiments with our pipeline were performed using a Spark installation with multiple computing nodes, providing a total of $256$ GB RAM of memory per node. 

\subsection{Dataset} 
\label{sec:dataset}
 %\textbf{Q: different resolution images}\\
The dimensions of the images in the dataset were selected to enable a fair comparison between the software using the hardware resources available to us. Smaller dimensions were omitted as they fell outside the scope of our study. %Nevertheless, we note that \PIXH's efficient memory utilization enables computations even for smaller images.

We generated for our experiments a dataset of $90$ astronomical images using the Astropy library \cite{astropy2022}. The creation of the image starts with the creation of an array of dimensions  $10,000 \times 10,000$, where all pixels are set to zero. Gaussian readout noise and sky background are added. Stars are then generated and added to the image. The details of the process are described in \cite{CCDguide}. To make the images more realistic, we set the number of objects within each image to approximately $340,000$. The resulting images with a precision of \emph{float32} were saved in \emph{.fits} format, resulting in a dataset size of about $36$ GB.

\begin{figure}[ht]
 \centering
  \includegraphics[width=5cm]{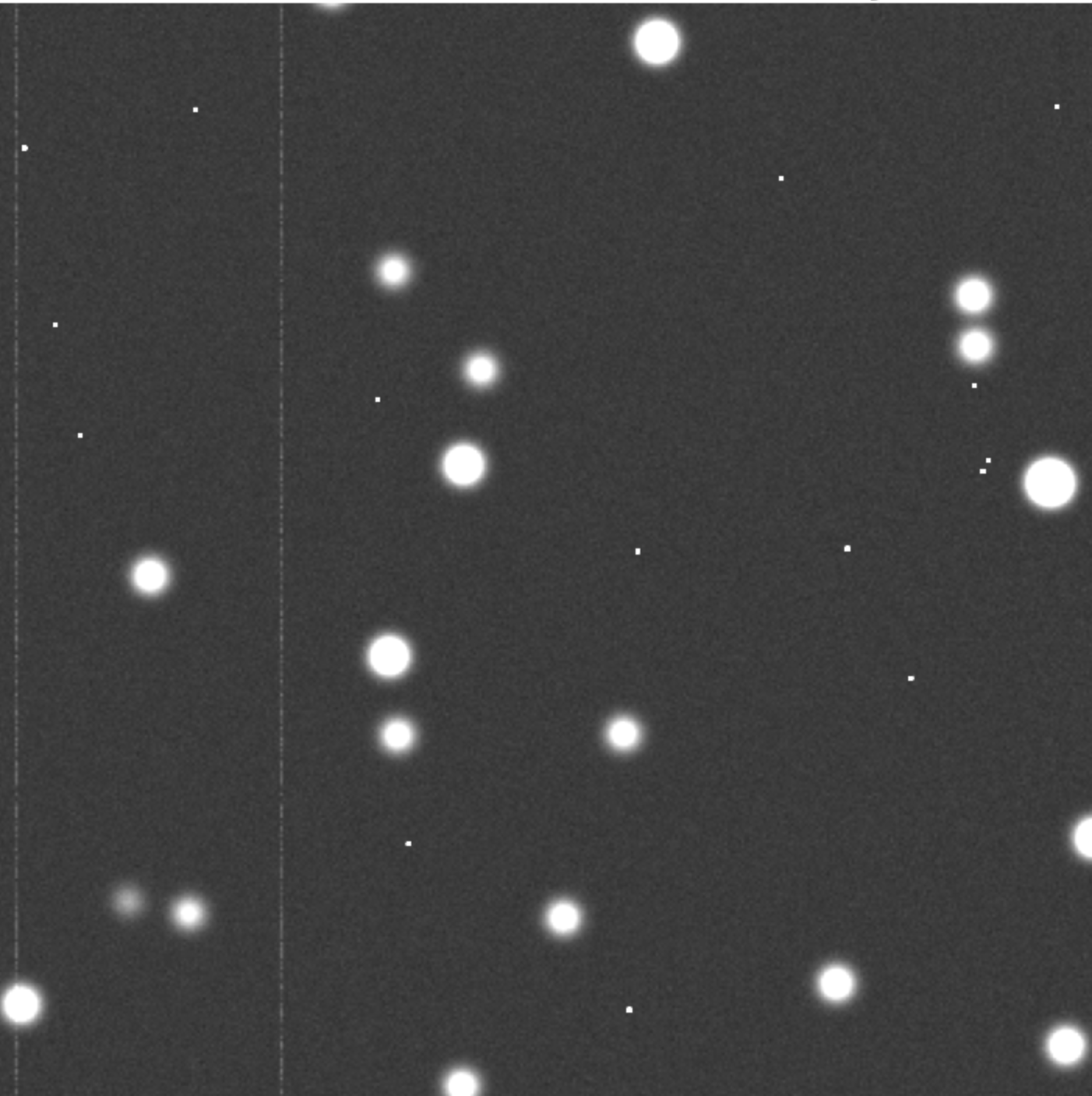}
  \caption{A cropped section of $500 \times 500$ pixels from an image in the dataset.}
  \label{fig:ex}
\end{figure}

%{\red We retrieved images from the Mikulski Archive for Space Telescopes (MAST) using the \emph{skyview.get\_images} function within the astroquery package in Python. The images were sourced from the SDSSu survey, featuring varied coordinates and pixel dimensions of $10,000\times10,000$. A total of $90$ images that satisfy the criteria for applying  Algorithm \ref{alg:pixh_alg}  were chosen and saved in the \emph{.fits} format. The dataset is approximately $36$ GB.}

\subsection{Performance Assessment of \PH}

In our first round of experiments, we analyzed the experimental performance of several different variants of our \PIXH algorithm in order to find the combination leading to the best performance.

\subsubsection{Variant 1: reducing images loading times}

In this experiment, we conducted a comparative analysis of the execution time required by two distinct approaches for handling the input dataset of images. The first approach requires the driver program to be responsible for loading the entire dataset in memory and distributing it to the computing nodes (i.e., \textbf{load\_driver}). The second approach (i.e., \textbf{load\_self}), detailed in Section \ref{subsubsec:variants}, allows the executors to load the images on their own.

This latter approach resulted to be significantly faster than the former (results not reported but available upon request). The observed efficiency of \textbf{load\_self} is mostly due to its ability to greatly reduce the communication overhead for the driver program, as well as to deeply decrease its memory requirements.

%For this reason, from this point on we will always refer to this optimized version of the algorithm for the remaining experiments.

\subsubsection{Variant 2: reducing the amount of pixels to process}
%{\red da rimuovere (?!?)}

In this experiment, we assessed the impact of different filtering rules on the performance of our algorithm. Namely, we evaluated how the number of background pixels removed by the filtering rule introduced in variant 2 (see Section \ref{subsubsec:variants})
affected the efficiency of the algorithm. We investigated four different scenarios:

\begin{itemize}
  \item \textbf{vanilla}: Baseline scenario: no filtering rule is applied.
  \item \textbf{filter\_std}: Standard filtering rule: all pixels below a predefined threshold are classified as background.
  \item \textbf{filter\_light}: More conservative filtering rule: only pixels  below 30\% of the predefined threshold are classified as background.
  \item \textbf{filter\_heavy}: More aggressive filtering rule: all pixels  up to 30\% above the predefined threshold are classified as background.
\end{itemize}

The results of this experiment, shown in Table \ref{tab:filtering_performance},  indicate that filtering rules can reduce the execution time of \PIXH by up to $10\%$, without any relevant degradation in the output quality. Instead, this optimization seems to have no significant effect on the performance of traditional \PH tools like \RIPSER.

\begin{table}[ht]
    \centering
    \begin{tabular}{lccc}
        \toprule
         & Dropped pixels ($\%$)& PixHom. time (min)& \RIPSER time (min) \\
        \midrule
        \textbf{vanilla} & - & $7.08 \pm 0.04$ & $9.28 \pm 0.04$ \\
        \textbf{filter\_light} & $4.19 \pm 1.98$ & $7.01 \pm 0.04$ & $9.28 \pm 0.03$ \\
        \textbf{filter\_std} & $4.68  \pm 2.25$ & $6.49 \pm 0.04$ & $9.29 \pm 0.04$ \\
        \textbf{filter\_heavy} & $5.08 \pm 2.48$ & $6.30 \pm 0.04$ & $9.29 \pm 0.04$ \\
        \bottomrule
        \end{tabular}
    \caption{Execution time comparison of \PIXH using different filtering threshold levels. For each test conducted on $10$ images of the dataset, the median and standard deviation are provided, along with the percentage of background pixels dropped. These times are compared with those of \RIPSER.}
    \label{tab:filtering_performance}
\end{table}

\subsubsection{Variant 3: improve the workload distribution.}

In this experiment, we measured the overall time required to analyze our entire dataset using the three partitioning strategies outlined in Section \ref{subsubsec:variants}: partitioning by the number of executors (\textbf{part\_executors}), partitioning by the number of images (\textbf{part\_images}) and partitioning according to the Longest Processing Time (LPT) rule ( \textbf{part\_LPT}).

We evaluated the performance of these strategies, including the impact of the number of available executors, under several different scenarios. We expected that the impact of these strategies would be more relevant when considering a large number of executors, as the average number of images to be processed per executor would decrease, leading to possible performance bottlenecks due to straggler executors (i.e., executors requiring a much longer time to complete than the other ones, causing delays in the overall execution). For this reason, we varied the number of executors used in our test from $2$ to $18$, while processing a constant number of images.

\begin{figure}[ht]
 \centering
  \includegraphics[width=9cm]{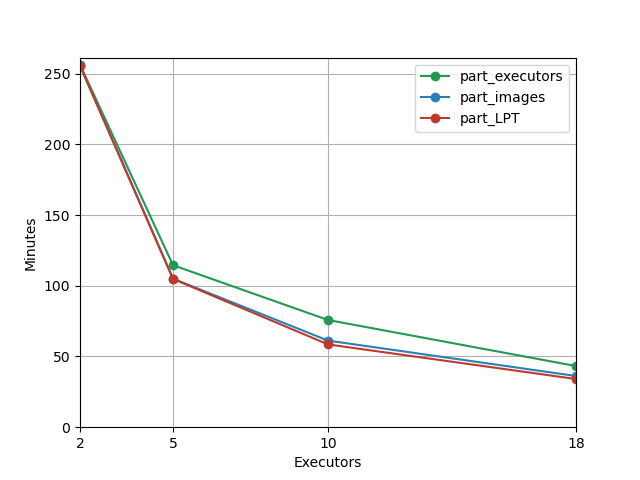}
  \caption{Comparative analysis of the execution time for analyzing the entire input dataset when using different partitioning strategies and an increasing number of executors.}
  \label{fig:exp}
\end{figure}

As expected, the results in Figure \ref{fig:exp} show that there is little difference between the three partitioning strategies when using a very small number of executors or a very large number of images. In this scenario, all strategies resulted in a very similar workload distribution. However, we observed a relevant change when increasing the number of executors and decreasing their average workload. Under these conditions, the \textbf{part\_executors}  strategy began to underperform, also because of its inability to reassign tasks to balance the workload. 

Moreover, the experiments proved the LPT-based strategy to require a lightweight computation (taking at most 20 seconds) and be slightly faster than the others, although the performance improvement was very small. On closer inspection, we found that the execution cost estimation, which is based on the number of background pixels, can be inaccurate, sometimes leading to suboptimal task assignments in the LPT strategy.

\label{sec:scalabilityandpartitioning}
\label{sec:partexec}

\subsubsection{Final configuration}
Based on the results of the experiments presented so far, we determined the most effective variants to use for the next experiments. The selected variants are:
\begin{itemize}
  \item Variant 1: we chose \textbf{load\_self} for its efficiency when loading input images.
  \item Variant 2: we chose \textbf{filter\_std} for its balanced approach to image filtering.
  \item Variant 3: we chose \textbf{part\_LPT} thanks to its improved performance in workload distribution.
\end{itemize}

\subsection{Experimental Comparative Analysis:  \RIPSER}

As described in Section \ref{sec:relatedwork}, \RIPSER is the current gold standard software for computing \PH filtrations across all dimensions. Thus, we used it in our experiments to evaluate both the quality of the output of our algorithm and its performance, when applied to the particular case of $0$-dimensional \PH computation.  

%Finally, we performed a qualitative comparison of the outputs of the two algorithms by using the bottleneck distance metric.

In our first experiment, we performed a qualitative comparison of the outputs from the two algorithms using the {\em bottleneck distance} (\cite{bottleneckdist}). It is an index measuring distances between two persistence diagrams as a minimal matching between them, allowing points to be matched with the diagonal $\Delta$, which is the set of all $(x,x) \in \mathbb{R}^2$. To this end and for visualization purposes, we selected a $50\times50$ pixels reference image, obtained by cropping one image chosen at random from our dataset.

\begin{figure}[ht]
 \centering
  \includegraphics[width=9cm]{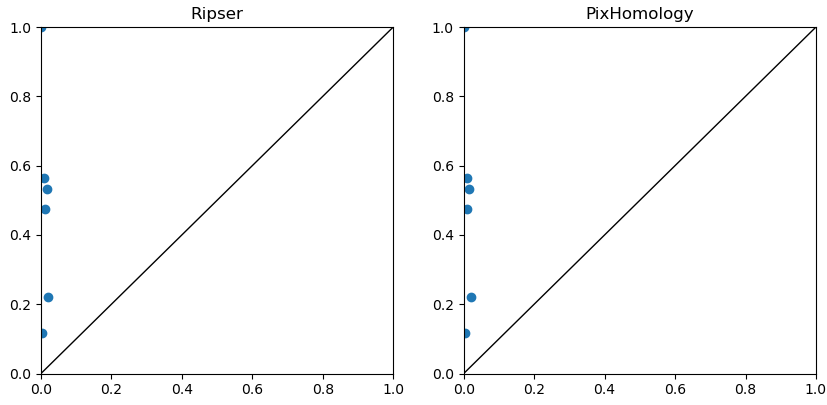}
  \caption{PD representation of the outputs from \RIPSER and \PIXH, when used to process a $50\times50$ pixels reference image obtained by cropping one image chosen at random from our dataset. The \pbirth and \pdeath points identified by the two algorithms are consistent. }
  \label{fig:comparison_diagram}
\end{figure}

As shown in Figure [\ref{fig:comparison_diagram}], the {\PD}s returned by the two algorithms are consistent, suggesting that the two methodologies produce similar results. To validate this outcome, we computed the bottleneck distances between the two diagrams, which resulted to be zero.
%
%\begin{definition}
%Let $\mathcal{M}$ be the space of all finite metric spaces. Let ${\rm dgm}(X,d_X)$ be the \PD corresponding to the \PH of $(X,d_X)$.  Let $\mathcal{D}=\{{\rm dgm}(X,d_X) \mid (X,d_X)\in\mathcal{M}\}$ be the space of \PD.  The {\em bottleneck distance} on $\mathcal{D}$ is given by
%$$
%d_{\infty}({\rm dgm}(X,d_X),\, {\rm dgm}(Y,d_Y))= \inf_\gamma\sup_{x\in{\rm dgm}(X,d_X)}\|x-\gamma(x)\|_\infty,
%$$
%where $\gamma$ is a matching of points between ${\rm dgm}(X, d_X)$ and ${\rm dgm}(Y, d_Y)$. 
%\end{definition}
%

\begin{figure}[ht]
 \centering
  \includegraphics[width=9cm]{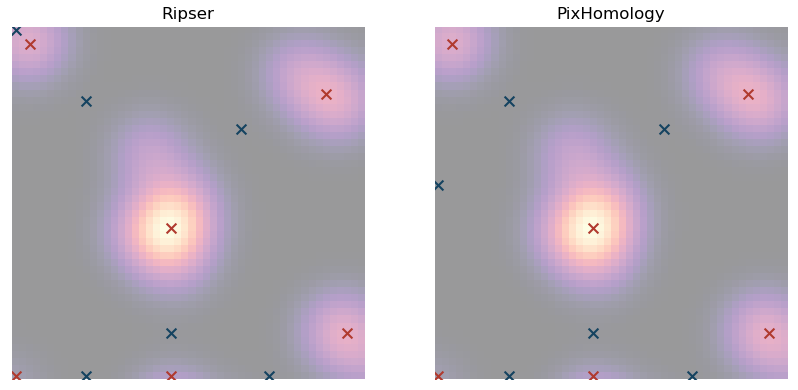}
  \caption{Pixel representation of the outputs from \RIPSER and \PIXH for a $50\times50$ pixels reference image obtained by cropping one image chosen at random from our dataset. The \pbirth and \pdeath identified on the filtered image reveal a discrepancy at the point of infinity between the two methods.}
  \label{fig:comparison_output}
\end{figure}

Figure \ref{fig:comparison_output} compares the pixel-based outputs of the two algorithms for the same image. Birth pixels are marked in red and death pixels in blue. The position of one point, specifically the point at infinity, differs between the two outputs. This is because \RIPSER does not return the positions of birth and death points in pixel coordinates, requiring an additional reconstruction step. This limitation can pose serious identification challenges, especially when multiple pixels share the same value in the image. In contrast, \PIXH returns the positions of birth and death points in pixel coordinates, thereby circumventing this issue.

Next, we compare the experimental performance of the two algorithms in terms of total execution time and overall memory usage. To ensure a fair comparison and due to the inability of \RIPSER to take advantage of distributed or multicore systems, we executed \PIXH using one single executor running on one single core.

We expect \RIPSER to perform very well on small images, while our algorithm is expected to handle better very large images. To assess this expectation, we designed our experiment to evaluate the performance of the two algorithms on our entire dataset, but using a random crop of each image, with a size varying from $20\times20$ to $10,000\times10,000$ pixels. 

\begin{figure}[ht]
 \centering
  \includegraphics[width=9cm]{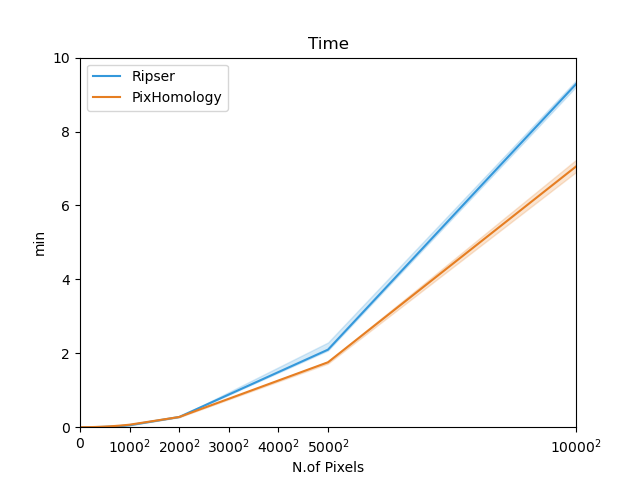}
  \caption{Execution time comparison between \RIPSER and \PIXH, for analyzing the images of our dataset,  using a random crop of each image, with a size varying from $20\times20$ to $10,000\times10,000$ pixels. \PIXH was executed on a single-core single-executor Spark distributed system.}
  \label{fig:comparison_time}
\end{figure}

\begin{figure}[ht]
 \centering
  \includegraphics[width=9cm]{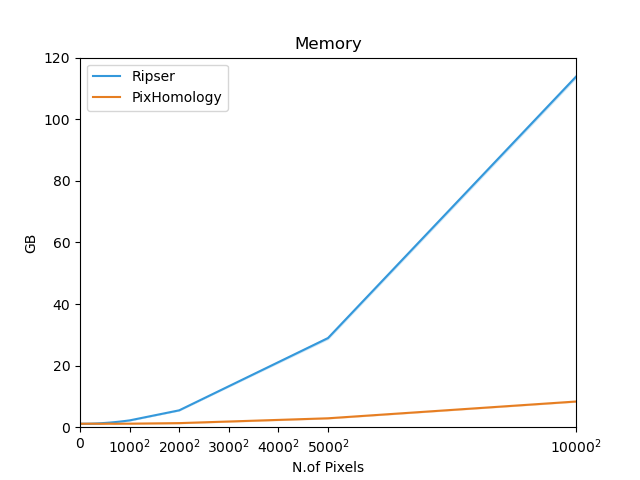}
  \caption{Maximum memory usage comparison of  \RIPSER and \PIXH for analyzing the images of our dataset,  using a random crop of each image, with a size varying from $20\times20$ to $10,000\times10,000$ pixels. \PIXH was executed on a single-core single-executor Spark distributed system.}
  \label{fig:comparison_memory}
\end{figure}

The resulting execution times are presented in Figure \ref{fig:comparison_time}. As expected, \RIPSER outperforms \PIXH on small patches. However, as the image size increases to $1,000\times1,000$ and beyond, \PIXH becomes much faster than  \RIPSER. The differences in memory usage between the two algorithms are even more relevant,  as shown in Figure [\ref{fig:comparison_memory}]. Notably, for  $10,000\times10,000$ pixel patches, \RIPSER requires approximately $112$ GB of memory because of the adjacency matrix computations. In contrast, \PIXH only requires around $8$ GB. 

We remark that the large memory requirements of  \RIPSER limit its applicability to the analysis of very large images and prevent running multiple instances of it concurrently on a single machine. Conversely, \PIXH's efficient memory usage makes it a more viable option for large-scale parallel processing on many core systems.

%These findings makes \PIXH as a formidable algorithm for parallel image processing.

%The substantial memory requirements of \RIPSER make it impractical to run multiple instances at the same time on a multi-core machine. On the contrary, \PIXH's efficient memory management makes it a more suitable option for simultaneous execution on such hardware.

\subsection{Experimental Comparative Analysis:  \DIPHA}
We were interested in assessing the performance of \PIXH when run as a distributed pipeline, in terms of execution times and scalability. For this purpose, we benchmarked its performance against  \DIPHA, the only existing distributed pipeline designed for large-scale \PH computation.

We recall that the strategy adopted by \DIPHA for the distributed \PH calculation implies that each computing unit of the distributed system analyzes a segment of the input image, with the size of this segment being inversely proportional to the total number of computing units. Consequently, when processing very large images using a few computing units, each unit requires a significant amount of memory to run.
This was a limiting factor in our experiments, as we were unable to execute \DIPHA with only two computing units because of the memory required per unit, due to its high memory requirements (approximately $9$ GB per unit).

\begin{figure}[ht]
 \centering
  \includegraphics[width=9cm]{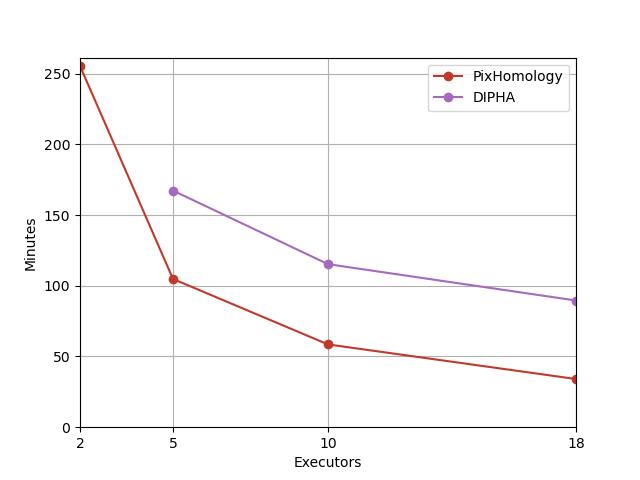}
  \caption{Execution time comparison between \DIPHA and \PIXH, for analyzing the images of our dataset, using an increasing number of computing units.}
  \label{fig:pixvsdipha}
\end{figure}

For this reason, the analysis of our entire dataset with \DIPHA was only possible when using a large number of computing units. In such a setting the results, available in Figure [\ref{fig:pixvsdipha}] show that, although both algorithms exhibit a very good scalability, \PIXH consistently outperforms \DIPHA in terms of execution time. 

%Furthermore, our algorithm has a memory footprint much smaller than the one of \DIPHA. 

\section{Conclusions and Future Directions}
\label{sec:conclusions}

In this paper, we presented \PIXH, a novel and efficient algorithm for computing \PH on large batches of images. We also presented a distributed version of \PIXH able to leverage the Apache Spark computing framework to concurrently process large batches of digital images. In addition, we also considered several variants of our distributed algorithm, for addressing performance bottlenecks identified during preliminary experimental evaluation. 

We compared the performance of our algorithm to that of the state-of-the-art algorithm in this field, finding that \PIXH is significantly faster and less memory-demanding for processing batches of large digital images. We also compared \PIXH to the only other distributed pipeline for PH computation available and again found that our algorithm is more favorable in terms of execution time and memory requirements.

Despite this progress, there is still room for improvement and expansion. Primarily, there is potential to enhance the algorithm's versatility for application across diverse image types. To address this, a possible solution involves introducing controlled noise into the image, mitigating the likelihood of neighboring pixels adopting identical values. Another avenue for exploration is the potential to distribute the analysis of a single image by partitioning the image processing workflow into smaller, more manageable sub-processes. This partitioning strategy could significantly enhance the scalability of our pipeline and further optimize memory utilization. A further problem occurs when the size of input images exceeds the amount of memory available to each executor (see Section \ref{subsubsec:variants}). Possible solutions to address this problem would be either to partition each images into independent parts to be processed by different executors or develop out-of-core algorithms for this purpose.

Additionally, comprehensive testing and evaluation of the pipeline's performance on other application domains, such as the analysis of diagnostic images, is essential. This would thoroughly assess the versatility and adaptability of the pipeline and ensure that it remains a robust solution across diverse scientific disciplines. 
Notably, \PH encounters challenges when applied to large-scale images, as evident in possible tasks like tumor segmentation from whole slide histology images  \cite{FastTumorSegmentation} and astronomical image segmentation \cite{tis}.
%{\red Thus, we are working to extend our analysis on real-world datasets to assess how \PIXH behaves  when processing images from different domains and with different resolution. Future experiments are planned to be performed on a broader HPC infrastructure to validate the scalability and robustness of our algorithm, as well as to explore additional optimization strategies.}

Finally, a research line that has still to be explored but has a potential for giving breakthroughing results, is about the possibility to employ machine learning techniques to solve the \PH problem. Indeed, such an approach may lead to results that may be, to some extent, less accurate than deterministic solutions but in a fraction of time.

\section{Funding}
Università di Roma - La
Sapienza Research Project 2021 “Caratterizzazione, sviluppo e sperimentazione di algoritmi efficienti”. INdAM – GNCS Project 2023 “Approcci computazionali per il
supporto alle decisioni nella Medicina di Precisione”. STILES, ``Strengthening the Italian leadership in ELT and SKA'', financed by the EU and approved by the MUR (Ministerial Decree no. 3264 of December 28th, 2021, ``Creation of new research infrastructures'') based on the NRRP - NextGenerationEU - Mission 4 ``Education and Research'', Component 2 ``From research to business'' - Investment 3.1, ``Fund for the realisation of an integrated system of research and innovation infrastructures''.

\bibliographystyle{abbrv}
\bibliography{references}  %%% Uncomment this line and comment out the ``thebibliography'' section below to use the external .bib file (using bibtex) .





% \end{thebibliography}
%\printbibliography

\end{document}